# The IBIS / ISGRI Source Location Accuracy

A. Gros[1] [a b], A. Goldwurm [a c], S. Soldi [a b], D. Götz [a b], I. Caballero [a b], F. Mattana [c], J. Zurita Heras [c]

[a] Service d'Astrophysique / IRFU / CEA – Saclay, 91191 Gif sur Yvette, France
[b] Astrophysique Instrumentation et Modelisation, CEA – Saclay, 91191 Gif sur Yvette, France
[c] AstroParticule et Cosmologie, 10 rue Alice Domon et Léonie Duquet, 75013 Paris, France

[1] E-mail : aleksandra.gros@cea.fr

We present here results on the source location accuracy of the INTEGRAL IBIS/ISGRI coded mask telescope, based on ten years of INTEGRAL data and on recent developments in the data analysis procedures. Data were selected and processed with the new Off-line Scientific Analysis pipeline (OSA10.0) that benefits from the most accurate background corrections, the most performing coding noise cleaning and sky reconstruction algorithms available. We obtained updated parameters for the evaluation of the point source location error from the source signal to noise ratio. These results are compared to previous estimates and to theoretical expectations. Also thanks to a new fitting procedure the typical error at 90% confidence level for a source at a signal to noise of 10 is now estimated to be 1.5 arcmin. Prospects for future analysis on the Point Spread Function fitting procedure and on the evaluation of residual biases are also presented. The new consolidated parameters describing the source location accuracy that will be derived in the near future using the whole INTEGRAL database, the new fitting technique and the bias correction, will be included in future versions of OSA.

---

[1] Speaker

## 1. Data and technique

In this study of the point source location accuracy of the imaging system of the IBIS / ISGRI telescope (Ubertini et al. 2003, Lebrun et al. 2003) onboard of the INTEGRAL satellite, we used most of the data collected from the Crab and Cyg X-1 observations carried out during the nearly ten years of INTEGRAL operations. They include more than 20000 images integrated over a science window (ScW), i.e. a period of data collection at a stable satellite attitude with typical exposure of 1800 s, in the four energy bands 20-40, 40-80, 80-150, 150-300 keV and processed with the most recent off line scientific analysis pipeline (OSA 10.0) (Goldwurm et al. 2003) delivered by the INTEGRAL Science Data Center (Courvoisier et al. 2003). Following the method described in Gros et al. (2003), the source excess in a deconvolved image is fitted with a function that approximates the final Point Spread Function (PSF). The angular offset of the derived position with respect to the catalogue source position is then measured. The employed PSF function is a bidimensional Gaussian with free width along both axes. Fig. 1 shows the typical profiles along the two axes in a reconstructed image of the source together with the fitted Gaussian model.

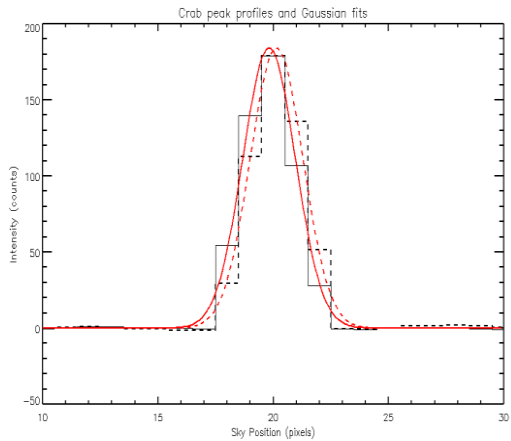
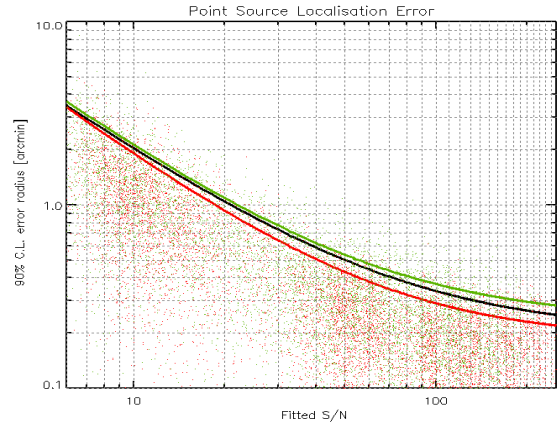

**Fig. 1** X and Y profiles of the Crab peak in a deconvolved image and of the fitted Gaussian.

**Fig. 2** Measured offsets from OSA 10 processed data (dots) and fitted 90% c.l. PSLE curves as a function of the SNR for the FcFoV (red), PcFoV (green) and the whole FoV (black).

## 2. Location Error as a function of the Source Signal to Noise Ratio

The offsets for a large number (N) of source measurements are plotted as a function of the source signal to noise ratio (SNR) derived from the fit in order to establish the Point Source Location Error (PSLE) at the 90% confidence level (c.l.) in arcminutes as a function of the SNR (Gros et al 2003). The data are binned in SNR to have 100 points per bin and the offset that include 90% of the events is computed.

A curve of the form

$$PSLE = a\ SNR^c + b$$

is then fitted to these 90% offsets. The measured offsets and the best fit curves are shown in Fig. 2 while derived parameters are reported in Table 1 for the offsets in the Fully coded Field of View (FcFoV), Partially coded FoV (PcFoV) and for all points (FoV). The errors on the parameters were determined using the bootstrap technique (Simpson et al. 1986).

**Table 1** Best fit parameters for the assumed law that describes the location accuracy at 90% c.l. as a function of SNR ($PSLE = a\ SNR^c + b$) for different parts of the FoV.

| FOV | N | a | b | c |
|---|---|---|---|---|
| FoV | 20402 | 24. ± 2 | 0.20 ± 0.02 | -1.12 ± 0.04 |
| FcFoV | 7803 | 28. ± 4 | 0.18 ± 0.02 | -1.21 ± 0.07 |
| PcFoV | 12605 | 24. ± 2 | 0.21 ± 0.02 | -1.09 ± 0.04 |

**3. Comparison with previous results**

Fig. 3 compares OSA 3 PSLE derived for the whole FoV (Gros et al. 2003) to later estimates for OSA 7 in the FcFoV and PcFoV (Scaringi et al. 2010), to the present OSA 10 results and to the best possible (theoretical) location error. Location errors decreased from OSA 3 to OSA 7 and from OSA 7 to OSA 10 processed data. The fit quality is given by the fraction of points below the fitted curve. Fig. 4 highlights this progress in terms of sky search area decrease defined as the relative percentage difference between two location areas corresponding to two error radii (Scaringi et al. 2010)

$$gain = 100 \cdot \frac{(Area_2 - Area_1)}{Area_1}\ ,\ Area \propto PSLE^2$$

An average gain of 20% in the FcFoV and 10% in the PcFoV from OSA 7 to OSA 10 is observed. For large signal to noise ratios, the negative influence of the Crab, due to its intrinsic location bias (Eckert et al. 2010), possibly due to the source extension, is seen. Therefore the analysis was also performed for the data that do not include the Crab.

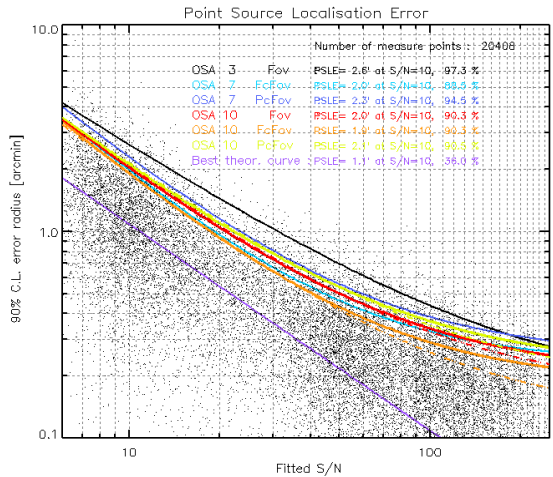 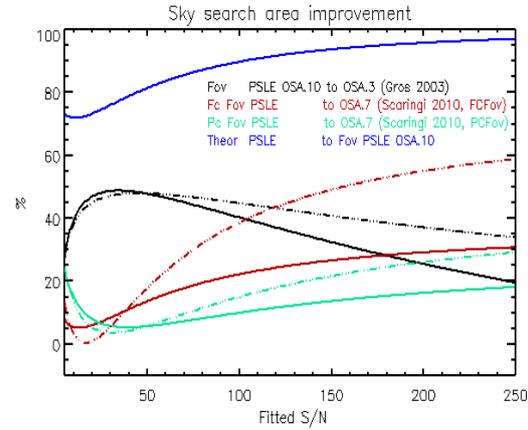

**Fig. 3**  PSLE curves for OSA 3 (black), OSA 7 FcFoV (light blue), OSA 7 PcFoV (dark blue), OSA 10 FoV (red), OSA 10 FcFoV (orange), OSA 10 PcFoV (yellow) and best theoretical PSLE (purple). Dashed lines are for the data that do not include Crab.

**Fig. 4**  Sky search area gain with OSA 10: OSA 10 to OSA 3 (black), OSA 10 to OSA 7 in FcFoV (red), same in PcFoV (green). The maximum (theoretical) gain (blue) is also indicated. Dashed lines are for the data that do not include the Crab.

## 4. Fitting with different methods

Different fit methods were tested in order to improve the source localisation accuracy. In this study the Crab data were excluded, in order to avoid the possible inherent shift and ~15000 data-points in the whole FoV were used, with their raw SNRs, instead of the fitted ones, to allow for a valid comparison between the different fit methods. First the width of the Gaussian function used in source peak fitting was fixed to its FcFoV expected value (Gros et al. 2003). This improved slightly the PSLE (Fig. 5). The relative sky search area decrease with respect to the default OSA 10 curve shows an average gain of ~10% (Fig. 6).

A more significant improvement was achieved when the $\chi^2$ of the fit was computed by weighting residuals with the image pixel significance. This enabled a gain in the sky search area of ~30% for weak sources. The new 90% c.l. PSLE at SNR=10 is 1.5 arcmin, significantly better than the 2.0 arcmin of the previously published curves (Scaringi et al. 2010). The coefficients for this PSLE are given in Table 2.

**Table 2**  Best fit parameters of the 90% c.l. PSLE as a function of the SNR for the whole FoV using a PSF fit where deviations are weighted by the pixels significance.

| N | a | b | c |
|---|---|---|---|
| 15206 | 17 ± 1 | 0.150 ± 0.006 | 1.10 ± 0.02 |

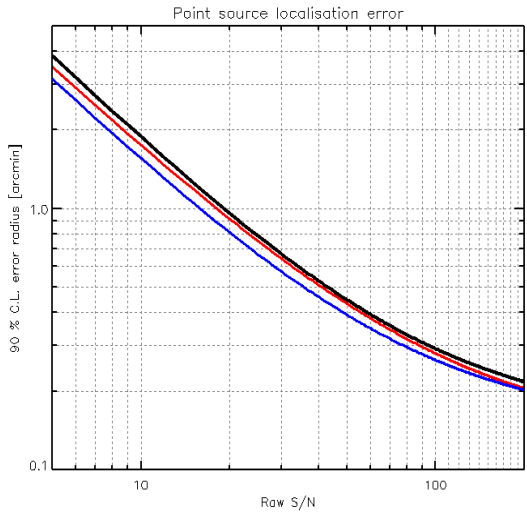 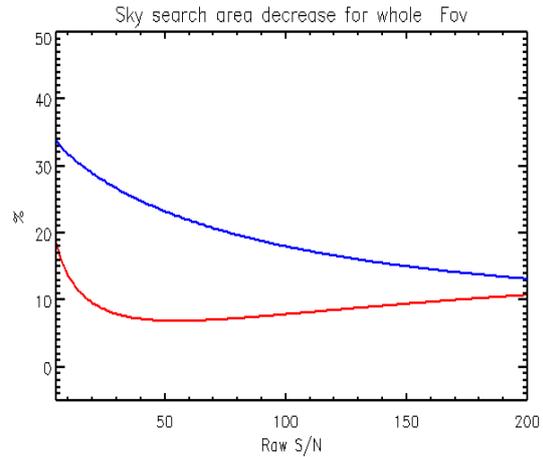

**Fig. 5** PSLE curves in the whole FoV fitted to the OSA 10 data vs. the raw SNR, for the standard fit with a free-width-Gaussian (black), with a fixed-width-Gaussian (red) and a fixed-width-Gaussian with weighted $\chi^2$ (blue).

**Fig. 6** Sky search area decrease with respect to OSA 10 default PSLE for a fixed Gaussian width (red) and for a fixed Gaussian width and weighted $\chi^2$ (blue).

## 5. Systematic shift

From the measured offset distribution one can estimate a possible systematic shift of the source from its catalogue position. Given the expected relation between offset and SNR, one should consider each given offset as a measure of the possible source shift with an error inversely proportional to the current source SNR associated with this offset measurement.

Note that this approach takes into account the observation conditions (time, source distance from axis, etc). Thus, the source shift should be calculated as the mean of offsets weighted by the inverse of the squared SNR (Bevington 2003). There is no need to bin the offset distribution as done in Eckert et al. (2010) since this makes the result of the fit dependent on the binning method. Note that the offset distribution is not necessarily a Gaussian one, as for the Cyg X-1 offset distribution shown in Fig. 7, where significant tails are clearly present.

## 6. Conclusions and perspectives

Improved parameters for the PSLE of the IBIS/ISGRI telescope have been provided using a large data set and the most recent OSA 10 analysis pipeline. We have also identified a better PSF approximation and fit method for the evaluation of the source location. We plan to further validate this technique by using the whole INTEGRAL data base in order to establish the most precise PSLE for the IBIS/ISGRI telescope. This work will naturally also give a more precise evaluation of the possible residual systematic shifts in source location and their dependence with observation parameters. The new consolidated fitting technique and the derived improved parameters will be implemented in the next OSA versions.

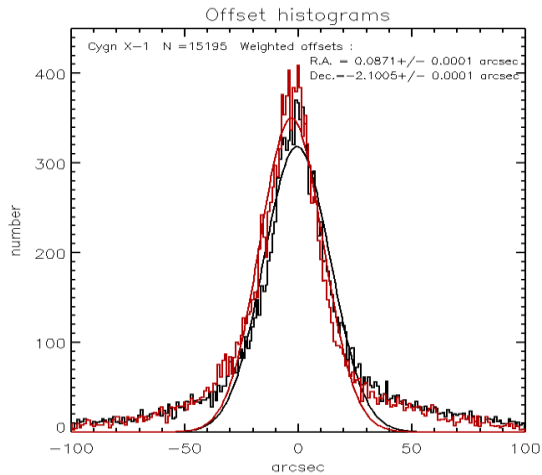

**Fig. 7** Distribution of the offsets for Cyg X-1 in R.A. (Black) and Dec. (red) compared to the fitted Gaussians.

**Acknowledgments**

The present work is based on observations with INTEGRAL, an ESA project with instruments and science data centre funded by ESA member states (especially the PI countries: Denmark, France, Germany, Italy, Switzerland, Spain, Czech Republic and Poland, and with the participation of Russia and the USA). ISGRI has been realized by CEA-Saclay/IRFU with the support of the Centre National d'Etudes Spatiales (CNES). SS, IC, FM and JZH acknowledge the CNES for financial support.